\documentclass[oldversion]{aa}
\usepackage{graphicx}
\usepackage{txfonts}
\usepackage{natbib}
\usepackage{mathrsfs}
\usepackage{multicol}
\usepackage{esvect}
\usepackage{calligra}
\usepackage{longtable}
\usepackage[T1]{fontenc}
\usepackage{url}

\bibpunct{(}{)}{;}{a}{}{,} 

\usepackage{siunitx}

\usepackage{hyperref}
\begin{document} 

\title{''Forbidden" stars in the eROSITA all-sky survey:\\
X-ray emission from very late-type giants}

\author{J.H.M.M. Schmitt\inst{1}, M. H\"unsch\inst{1},  P.C. Schneider\inst{1}, S. Freund\inst{2}, S. Czesla\inst{3}, J. Robrade\inst{1}, A. Schwope\inst{4}}

\institute{Hamburger Sternwarte Universit\"at Hamburg, Gojenbergsweg 112, 21029 Hamburg, Germany
\and
Max-Planck-Institut f\"ur Extraterrestrische Physik, Gie\ss enbachstra\ss e 1, 85748 Garching bei M\"unchen
\and
Th\"uringer Landessternwarte Karl-Schwarzschild- Observatorium, Sternwarte 5, 07778 Tautenburg
\and
Leibniz-Institut f\"ur Astrophysik, An der Sternwarte 16, 14482 Potsdam}

\date{Received -; accepted -}

 
\abstract
{We present the results of the first X-ray all-sky survey (eRASS1) performed by the eROSITA instrument onboard the Spectrum-Roentgen-Gamma (SRG) mission 
on X-ray emitting red giants and supergiants.      Focussing on stars positioned at high galactic latitudes above 20$^{\circ}$, we construct a complete sample of such objects using the {\it Gaia} DR3 catalog and identify a sample 96 stars appearing as bona fide entries in the eRASS1 source catalog.   Restricting again the sample to objects nearer than 1300~pc and eliminating all catalog entries which are due to optical contamination, we end up with a sample of 16 genuine red giant/supergiant X-ray sources, which represent -- with the exception of one source (CL~Hyi) -- new X-ray detections.  We furthermore present a low SNR X-ray spectrum of the nearby low activity giant Arcturus obtained from a pointed observation with the XMM-Newton satellite and give a detailed account of our data analysis. We show that Arcturus-like X-ray emission cannot be the explanation for the X-ray emissions observed by eROSITA and provide a discussion of the possible nature of the
detected X-ray sources.} 

\mail{J.H.M.M. Schmitt, jschmitt@hs.uni-hamburg.de}
\titlerunning{eROSITA detected red supergiants}

\keywords{Stars: activity; Stars: X-rays; Stars: late-type}

\maketitle


\section{Introduction}
\label{sec_intro}
In flux-limited, large area soft X-ray surveys like those carried out by the {\it Einstein} and
ROSAT observatories (\cite{1991stocke}, \cite{2022freund})
``normal'', i.e., coronal, stars constitute a significant fraction of the overall source population and
are responsible  typically for about a 
fourth to a third of the detected X-ray source population; in this context we refer to ``normal'' stars as
stars for which accretion processes do not play any significant role for the generation of X-ray emission.
The vast majority of these X-ray detected objects are late-type stars on or near 
the main sequence
with rather deep outer convection zones, where -- presumably -- dynamo-related magnetic activity processes
similar to those observed on the Sun take place, but at levels usually much higher than the
emission levels encountered in the Sun.  Other ``normal'' stars would be early-type stars
of spectral type O or B and (non-accreting) white dwarfs, however, these source types account
only for a very small fraction of the overall stellar X-ray source population detected, for example, by eROSITA.

In stars with outer convection zones coronal X-ray emission is essentially ubiquitous as
demonstrated by volume-limited X-ray surveys of late-type (main-sequence) stars in
the immediate solar vicinity where one achieves X-ray detection rates of almost
100\% \citep{1997schmitt}.  This finding lends strong support to the view that magnetic activity and X-ray emission is indeed
a very typical characteristics of such stars and directly leads to the
rotation-age-activity paradigm: stars are born as rapid rotators with ensuing high levels
of magnetic activity.    Magnetic braking then leads to a slow-down of stellar rotation
which in turn leads to a reduction of activity, eventually to the levels known from the Sun,
which is quite inactive compared to a typical stellar source detected in
an X-ray survey.

This paradigm applies to stars on or near the main-sequence, and the question arises whether it also
applies to giants of late spectral type.  For such stars the  picture, which we briefly
sketch in the following, is more complicated.
Evolved giants have an internal structure very much different from their dwarf (main-sequence) siblings. 
Stars of low to intermediate mass leave the main-sequence once the core hydrogen (H) is exhausted. 
In such stars H-burning continues in a shell around an inert helium core, while the outer regions
expand, so that the stars become sub-giants and eventually red giant branch (RGB) stars. 
With the star moving up the RGB, the outer convective envelope becomes deeper and deeper  
and shell energy production increases.  Once at the tip of the RGB, the core mass 
is sufficiently large and the core temperature sufficiently high to start He-burning.  He burning starts
with a He-flash, and finally the stars arrive on the horizontal branch, i.e., the He burning main sequence. 
Stars of solar mass spend a comparatively long time in that stage and are therefore prominently seen in 
many surveys, including X-ray surveys. Eventually, all of the core He is transformed into carbon and oxygen,
He burning in the core ceases, and the star travels up the asymptotic giant branch (AGB). 

What about activity phenomena in such giant stars ?  As described above, late type giants do have deep convection zones, which 
is one of the prerequisites for stellar activity, yet the rotational velocities of such stars are usually quite low, as 
they were already slow rotators 
on the main-sequence or were slowed down during their evolution towards the giant phase, depending on the initial mass.  However, there are exceptions 
to this ``rule'':  in close binary systems tidal forces can maintain large rotational velocities and,
for example, the short period RS CVn binaries, which are composed of two giants of late spectral type, are among 
the brightest coronal sources.   The very few known fast rotating single giants are either situated in the 
Hertzsprung gap or may have formed out of coalesced binaries of the W UMa type \citep{1981bopp}.

Extensive studies of X-ray emission among giant stars have been carried out 
with the {\it Einstein Observatory} \citep{1981ayres} and 
ROSAT (\cite{1991haisch}, \cite{1992haisch}, \cite{1996huensch}, \cite{schroeder1998}, \cite{1998huenscha}). 
These studies  have indicated at least a strong decrease in X-ray emission 
towards later spectral types with a virtual extinction of X-ray emission near spectral type K3. This decrease in X-ray emission
goes along with a corresponding decrease in UV emission lines, indicating the presence of hot transition regions.
These findings have led to the concept of a so-called X-ray dividing line (XDL)  in the HR diagram:  giants on the ``X-ray side'' of the XDL have hot coronae, while giants on the other side have massive cool winds as first realized by \cite{1979linsky} on the basis of IUE spectra,
and the non-detection of X-rays to the right of the XDL has been attributed to the
absence of hot plasma in the low gravity atmospheres of these stars. 

No generally accepted physical interpretation of the XDL has been put forward.
\cite{huensch1996} have investigated the X-ray emission of late-type
giants in relation to their masses and evolutionary status and suggested that the XDL does actually
not reflect the decrease of stellar activity in the course of stellar evolution, yet they propose the XDL
to lie approximately parallel to the RGB, and it seems to be a consequence of lower stellar masses and longer evolutionary 
timescales. Hence the virtual absence of X-ray emission among giants later than K3 is more related to their rareness, their 
generally higher luminosities and larger distances, yielding less sensitive detection limits in X-ray surveys.
On the other hand, \cite{1991rosner} propose that the surface magnetic field configuration changes take place
as one moves along the giant and supergiant tracks in the Hertzsprung-Russell diagram and that these changes
are responsible for the appearance of the XDL.   Finally, there is the concept of ``buried coronae'' proposed, for example, 
by \cite{2004ayres}, who argues that the chromospheres of low-gravity giants are much thicker than those of
stars on the main-sequence.   Therefore any hot emission zones above the photospheres would be buried -- at least
partially -- by large columns of chromospheric material and so severely reduce the observable X-ray emission.

A prototypical example for such a ``buried corona'' is -- according to \cite{2018ayres} -- the nearby giant Arcturus (spectral type K1.5III, d =  11.3~pc), 
a low-mass Pop~II star of rather high age.
\cite{2018ayres} reports an X-ray detection at a level of $L_X \approx$ 3 $\times$ 10$^{25}$ erg/s obtained from a 
98~ksec exposure with the {\it Chandra} HRC;  this value is almost two orders of magnitude below solar
levels, and makes -- if typical -- such stars essentially undetectable if located at larger distances.   In terms of fractional
X-ray luminosity,  \cite{2018ayres} determined a value $L_X/L_{bol} \approx$ 5 $\times$ 10$^{-11}$, a value almost 5000 times
smaller than the corresponding value for the already rather X-ray dim Sun.
Unfortunately, the {\it Chandra} HRC does not provide spectral resolution at X-ray wavelengths, so 
no information on the energy distribution of the Arcturus' X-ray emission is available. 

However, more sensitive X-ray observations by ROSAT revealed X-ray emission of various giants of later spectral types,
which seemed to defy the concept of an XDL.    One such class are the hybrid stars, which are giants that
show simultaneously both the signatures of hot coronae {\bf and} cool winds.  \cite{1992reimers} and  \cite{1996reimers} demonstrated
the ubiquity of X-ray emission among hybrid stars based on pointed observations with the ROSAT PSPC.
\cite{2005ayres} challenged some of the detections reported by Reimers and collaborators, arguing that nearby 
companions were actually responsible for the detected X-ray emission, however, a deeper pointing with {\it Chandra}
showed at least for one of the disputed cases, i.e., $\gamma$ Dra, the clear presence of an X-ray source exactly at the
expected position \citep{2006ayres}.

Even farther on the cool side of the XDL are the M-type giants.
\cite{1998huenschb} used the ROSAT all-sky survey data to search for X-ray emission from ``bona fide'' M-type giants
by matching a sample of 482 such stars extracted from the Bright Star Catalog and found eleven positional coincidences.
Among those, \cite{1998huenschb} expect three spurious coincidences, and
in four cases the X-ray emission was very likely related to the symbiotic nature of the stars, i.e., these stars have known companions, while  the remaining four cases were identified as bona fide candidates for possibly intrinsic X-ray emission from such stars, which may of course be hitherto unknown binaries such
as symbiotic stars (SySts).  SySts are binary systems consisting of a red giant orbited by
a sufficiently hot companion, typically a white dwarf or neutron star; for a recent review of SySts we refer
to \cite{2019munari}.  High temperature material in such systems may be produced either by accretion or by
nuclear burning processes on the surface of the white dwarf.   Such system are naturally prone to produce outbursts
and various kinds of photometric variability.

Among the M-type giants, a large fraction are on the asymptotic giant branch (AGB), i.e., stars 
that have finished  their helium burning
phase, have an inert core consisting of carbon and oxygen and have thus considerably advanced 
in their overall stellar evolution.   Located far away from the dividing line, 
such stars are not really expected to intrinsically emit X-rays, yet some recent studies suggest that 
such stars can indeed be X-ray sources: \cite{2015sahai} carried out a pilot study and detected, using data from the XMM-Newton and {\it Chandra}
Observatories, X-ray emission from three (out of six) AGB stars preselected from an FUV excess.  \cite{2021ortiz} crossmatched
the XMM-Newton 4XMM-DR9 catalog of X-ray sources with a list of pre-selected AGB stars and ended up with a
sample of 17 reliable detections, six of them being new detections, while the remaining eleven objects had previously been
reported as X-ray sources detected with ROSAT, XMM-Newton  or {\it Chandra}.   The X-ray luminosities of these
objects are typically in the range 10$^{30}$ - 10$^{31}$ erg/s, i.e., far larger than ``typical'' coronal X-ray luminosities, and 
just like \cite{1998huenschb} these authors suggest that the underlying stars are SySts.

Thus binarity may indeed play an important role not only for our understanding of X-ray emission from those objects, but also
for an understanding of chemical peculiarities of some objects.   For example, \cite{2019jorissen} report the results of
a long term radial-velocity monitoring campaign of a few dozen barium stars, i.e., a class of red giants of spectral type G or K
with strong absorption lines of barium.   Orbital motion is detected in all systems with periods ranging from two years to hundreds of years,
and further, the masses of the companion white dwarfs are shown to be in the range 0.5 - 0.9 $M_{\sun}$.     Given that the
bright components in these systems are giants of substantial size, it is very clear that any periodicity due to a fainter
companion must occur on rather long time scales, requiring some stamina for optical followup.

Thus, the question to what extent very cool giants possess coronal plasma or not is still not fully clarified, and furthermore,
the role of binarity w.r.t the activity properties of these stars is also open.    In the mean time new data sets in the optical and X-ray range
have become available, which allow us to re-address these issues.   The eROSITA  ({\bf e}xtended {\bf RO}entgen {\bf S}urvey with an 
{\bf I}maging {\bf T}elescope {\bf A}rray) all-sky survey \citep{2021predehl} 
is far more sensitive than the ``old'' ROSAT all-sky survey and has a far better positional accuracy.
Furthermore with the {\it Gaia} DR3 release of the data from the {\it Gaia} mission \citep{2016gaia}
a very deep survey of the optical sky including parallaxes has become available.   With these data it is possible to construct an essentially
complete sample of red giants and supergiants in the solar ``neighborhood'' and conduct an unbiased search for X-ray emission from such objects.
And finally, a deep XMM-Newton pointing on Arcturus has become available which allows us to study the spectral properties of
a giant star very close to the dividing line.

The plan of our paper is therefore as follows:  we briefly discuss -- in Section \ref{sec_data} -- the new data used in our analysis and in particular the construction of a
complete sample.  In Section \ref{sec_res} we present our results: we discuss our identification procedures, the color-magnitude diagram
for the selected sources, the optical contamination problem, and our final candidate sample.  We furthermore present our analysis of a deep pointing on the nearby low-activity giant Arcturus and derive basic spectral properties of this source.  Finally, in section \ref{sec_discuss} we put everything together and present an in-depth discussion of our results.

\begingroup
\tiny
\onecolumn
\begin{center}
\begin{longtable}{   r   r  r   r   r   r   r   r r   c   c   }
\caption{\label{tab1} eRASS1 X-ray sources identified with very red giant stars; column 1 provides the Gaia DR3 identification number, columns 2 and 3 right 
ascension and declination of the optical counter part, columns 3 and 4 the Gaia apparent magnitude and BP-RP color, column 5 lists the
distance (in pc) as computed from the Gaia parallax, and column 6 provides a flag whether we consider the X-ray source to be caused
by optical contamination ("Y") or to be genuine ("N"); the asterisk in the last column indicates, whether the object is considered
as ``good'' candidate; see text for details.}
\\

\hline
\hline
DR3 number & RA   & Dec   & g & BP-RP & Distance & Detection & Match distance & Coronal & Contamination & cand \\
	\           & (deg) & (deg) &   &             & (pc)         &  likelihood &(arcsec)             & probability & flag & \ \\
\hline
\hline
\hline
4689639301195027840 & 6.0632 & -72.0768 & 10.91 & 2.63 & 4393.7 & 4.5 & 0.1 & N &   \cr
4710091076458784768 & 13.4085 & -62.8713 & 4.36 & 2.87 & 207.2 & 6.9 & 0.7 & Y &   \cr
4644856055150677632 & 32.9536 & -71.4841 & 6.78 & 4.31 & 645.9 & 5.5 & 0.96 & N & * \cr
4696456475143514368 & 36.36 & -66.494 & 4.98 & 2.95 & 224.7 & 2.2 & 0.95 & Y &   \cr
4693488382160219008 & 36.945 & -69.5238 & 4.93 & 3.84 & 240.0 & 2.4 & 0.96 & Y &   \cr
4738343783649552896 & 36.9477 & -57.6131 & 9.87 & 2.67 & 2923.9 & 4.1 & 0.86 & N &   \cr
5158832592439126272 & 41.983 & -12.4609 & 4.87 & 3.37 & 328.2 & 6.1 & 0.93 & Y &   \cr
4748477123328077696 & 43.4708 & -49.8895 & 6.12 & 5.6 & 231.3 & 1.6 & 0.98 & N & * \cr
5159194331764625920 & 44.7437 & -13.3438 & 7.47 & 2.64 & 769.7 & 1.3 & 0.93 & N & * \cr
4727337637895749888 & 48.1383 & -57.3215 & 4.54 & 2.52 & 460.5 & 2.2 & 0.98 & Y &   \cr
5100253911446411264 & 50.9019 & -19.8836 & 8.22 & 3.86 & 1291.7 & 3.3 & 0.97 & N & * \cr
4723045075781077760 & 52.1955 & -59.9767 & 8.22 & 2.64 & 1275.2 & 7.7 & 0.75 & N & * \cr
4732182395365306240 & 54.4358 & -55.3963 & 5.96 & 4.27 & 418.8 & 5.1 & 0.7 & Y &   \cr
4667961536094250752 & 56.7572 & -66.7707 & 8.82 & 2.75 & 1629.6 & 9.8 & 0.11 & N &   \cr
4829019511158220928 & 57.3994 & -52.0798 & 5.49 & 3.65 & 391.3 & 1.2 & 0.96 & Y &   \cr
4616109220563904768 & 62.3983 & -81.8548 & 8.09 & 3.75 & 324.9 & 0.5 & 0.98 & N & * \cr
4675293354507076608 & 65.4269 & -63.6584 & 8.01 & 3.82 & 910.3 & 11.2 & 0.0 & N &   \cr
4869021668323745792 & 66.3249 & -36.2088 & 8.43 & 3.01 & 1024.1 & 10.5 & 0.17 & N &   \cr
4663790332579168000 & 75.6147 & -65.1508 & 9.96 & 2.65 & 2145.7 & 9.8 & 0.61 & N &   \cr
3228743421412376704 & 76.3489 & 1.1776 & 4.58 & 2.97 & 490.2 & 2.6 & 0.99 & Y &   \cr
2989528414633822464 & 77.8454 & -11.8488 & 3.63 & 3.68 & 164.5 & 0.5 & 0.99 & Y &   \cr
4772353701224784640 & 81.9385 & -51.1776 & 8.46 & 2.96 & 1180.4 & 5.3 & 0.6 & N & * \cr
4660742383288287232 & 83.1331 & -65.8255 & 7.71 & 3.73 & 698.7 & 8.3 & 0.31 & N &   \cr
4650354712809914624 & 83.6866 & -73.7411 & 4.63 & 2.63 & 331.4 & 2.1 & 0.97 & Y &   \cr
4620697620025724544 & 88.5976 & -84.0563 & 7.86 & 3.01 & 792.9 & 1.5 & 0.93 & N & * \cr
5494400036544339072 & 90.539 & -60.0968 & 5.14 & 2.73 & 259.2 & 5.5 & 0.9 & Y &   \cr
5571300189028395008 & 91.0939 & -43.2978 & 9.61 & 4.68 & 2788.9 & 1.7 & 0.93 & N &   \cr
2913152694837288192 & 91.4398 & -24.1956 & 4.46 & 3.86 & 256.6 & 1.2 & 0.94 & Y &   \cr
5278871052851005184 & 96.7267 & -70.2638 & 11.81 & 2.58 & 5217.8 & 11.3 & 0.0 & N &   \cr
5283399421552134784 & 97.7546 & -66.8706 & 5.23 & 3.49 & 367.1 & 2.8 & 0.96 & Y &   \cr
5265304900428370560 & 99.7651 & -73.1979 & 10.8 & 3.05 & 3082.1 & 6.1 & 0.57 & N &   \cr
5262392225406506880 & 102.1676 & -73.2597 & 9.55 & 2.61 & 1867.6 & 8.2 & 0.0 & N &   \cr
893132234387415808 & 110.6175 & 32.4163 & 9.27 & 3.08 & 1196.1 & 4.6 & 0.88 & N & * \cr
5209380199986781184 & 127.6213 & -77.784 & 9.28 & 2.5 & 1776.7 & 3.4 & 0.77 & N &   \cr
5193680960648176768 & 128.1013 & -83.5025 & 8.53 & 3.79 & 1022.7 & 3.7 & 0.91 & N & * \cr
5761301055347596800 & 130.4596 & -5.6108 & 6.66 & 2.63 & 527.1 & 3.2 & 0.71 & N & * \cr
5209126861340475776 & 132.5494 & -78.7399 & 11.25 & 2.53 & 4969.1 & 8.0 & 0.97 & N &   \cr
612253364778494592 & 134.7945 & 18.1346 & 5.17 & 2.63 & 221.8 & 9.0 & 0.84 & Y &   \cr
699870869414974336 & 137.6616 & 30.963 & 3.37 & 3.95 & 150.4 & 1.3 & 1.0 & Y &   \cr
3842342137447011200 & 140.1529 & 0.1816 & 5.34 & 2.78 & 330.2 & 5.2 & 0.84 & Y &   \cr
645414291872801536 & 142.7756 & 25.0471 & 6.59 & 2.76 & 768.2 & 1.0 & 0.98 & N & * \cr
5195956296883496832 & 143.7215 & -80.1936 & 8.25 & 2.58 & 1041.0 & 4.7 & 0.78 & N & * \cr
794754943320201600 & 146.3929 & 34.5119 & 6.98 & 6.31 & 290.2 & 3.6 & 0.99 & N & * \cr
3827758180856268032 & 147.0311 & -0.862 & 9.65 & 3.87 & 1731.9 & 5.0 & 0.79 & N &   \cr
3550624862030637568 & 162.9052 & -21.2501 & 5.92 & 3.54 & 433.0 & 6.2 & 0.84 & Y &   \cr
3864344017954021120 & 164.006 & 6.1853 & 3.83 & 3.35 & 114.3 & 2.2 & 0.98 & Y &   \cr
3553091380145741440 & 165.1409 & -18.3249 & 5.05 & 4.99 & 212.6 & 1.0 & 0.94 & Y &   \cr
762126970124609408 & 167.3293 & 36.3093 & 4.43 & 2.78 & 131.6 & 4.0 & 0.98 & Y &   \cr
3910931906171050496 & 174.615 & 8.1343 & 3.91 & 2.78 & 179.7 & 2.1 & 0.99 & Y &   \cr
3491583497096278144 & 176.5414 & -24.8736 & 6.24 & 2.85 & 305.0 & 2.2 & 0.96 & N & * \cr
3463822649562256256 & 177.1634 & -35.987 & 4.16 & 4.05 & 212.5 & 1.5 & 0.99 & Y &   \cr
3490322803937287680 & 177.1877 & -26.7498 & 3.75 & 2.68 & 143.3 & 0.6 & 1.0 & Y &   \cr
3462933174719143424 & 183.3037 & -34.1252 & 4.43 & 3.48 & 437.3 & 0.9 & 0.98 & Y &   \cr
3707747654915318144 & 187.5874 & 4.4163 & 4.79 & 4.56 & 222.3 & 0.2 & 1.0 & Y &   \cr
3525277171936343680 & 192.4459 & -15.0789 & 5.49 & 3.15 & 425.8 & 5.3 & 0.93 & Y &   \cr
3944674165681089408 & 196.5943 & 22.616 & 4.05 & 2.96 & 183.0 & 2.1 & 0.99 & Y &   \cr
6186564871338685056 & 197.4043 & -27.4571 & 9.7 & 2.91 & 1871.0 & 5.6 & 0.77 & N &   \cr
3684575344281793920 & 198.5181 & -2.807 & 4.07 & 4.7 & 180.1 & 0.3 & 1.0 & Y &   \cr
6195030801635544704 & 202.428 & -23.2813 & 3.15 & 4.51 & 148.5 & 1.6 & 0.99 & Y &   \cr
3657349787109936896 & 208.0057 & -3.4802 & 8.41 & 2.67 & 852.5 & 11.6 & 0.17 & N &   \cr
6173329362681684736 & 209.43 & -31.0698 & 8.1 & 5.3 & 772.1 & 3.5 & 0.88 & N & * \cr
3674236602085548928 & 212.4996 & 7.3424 & 8.05 & 2.79 & 1185.7 & 4.0 & 0.96 & N & * \cr
6119285205089348608 & 214.9348 & -36.8583 & 4.87 & 3.9 & 175.0 & 3.4 & 1.0 & Y &   \cr
6323351986214357888 & 218.8267 & -13.4875 & 9.33 & 3.49 & 1895.6 & 8.5 & 0.5 & N &   \cr
6216707467220360448 & 219.949 & -32.0345 & 9.38 & 3.12 & 1722.9 & 2.3 & 0.98 & N &   \cr
6201910274096330368 & 222.1584 & -36.6349 & 4.12 & 3.29 & 167.9 & 0.6 & 0.97 & Y &   \cr
6313070693500197760 & 224.444 & -12.4377 & 5.07 & 3.9 & 338.5 & 5.2 & 0.9 & Y &   \cr
6207812177634139008 & 231.0517 & -31.7852 & 9.03 & 3.18 & 2097.5 & 9.8 & 0.27 & N &   \cr
6253880512249119104 & 235.1531 & -19.8685 & 10.74 & 3.64 & 1956.2 & 3.1 & 0.96 & N &   \cr
4395839245812332416 & 236.1152 & -8.4853 & 9.74 & 3.14 & 2928.5 & 7.1 & 0.8 & N &   \cr
5774447400486874112 & 262.8643 & -80.8593 & 4.7 & 2.61 & 241.1 & 10.8 & 0.24 & Y &   \cr
6738529604088911104 & 290.9611 & -39.9014 & 9.71 & 3.16 & 1205.3 & 6.1 & 0.6 & N & * \cr
6713213967334901376 & 290.9626 & -42.9361 & 7.49 & 2.51 & 807.0 & 7.0 & 0.8 & N & * \cr
6447070889998485376 & 298.8083 & -59.1958 & 4.38 & 5.12 & 174.1 & 3.2 & 0.99 & Y &   \cr
6444391655040244992 & 300.4366 & -59.376 & 3.14 & 3.12 & 169.2 & 1.6 & 0.98 & Y &   \cr
6443377905319179264 & 302.9414 & -59.9369 & 5.04 & 4.9 & 300.6 & 3.6 & 0.96 & Y &   \cr
6474933025745450880 & 309.4673 & -50.8643 & 11.35 & 2.91 & 4720.0 & 11.9 & 0.07 & N &   \cr
6342796540114659200 & 320.4733 & -86.4653 & 7.55 & 2.56 & 1045.1 & 5.6 & 0.76 & N & * \cr
6396597121531742336 & 322.1882 & -69.5056 & 3.76 & 3.2 & 127.6 & 2.6 & 0.94 & Y &   \cr
6459876523969640704 & 322.7516 & -56.8847 & 7.01 & 2.64 & 566.3 & 11.1 & 0.26 & N &   \cr
6351709524967494528 & 335.0085 & -80.4399 & 3.32 & 2.94 & 89.9 & 0.5 & 0.99 & Y &   \cr
6518817665843312000 & 335.6844 & -45.948 & 3.59 & 4.27 & 161.7 & 1.2 & 0.98 & Y &   \cr
6493430320313329920 & 344.2749 & -57.4014 & 4.9 & 3.9 & 230.2 & 4.5 & 0.96 & Y &   \cr
\end{longtable}
\end{center}
\endgroup

\twocolumn
\section{Data and data analysis}
\label{sec_data}

\subsection{SRG data}

For our study we used data from the
eROSITA instrument on board
the Russian-German Spectrum-Roentgen-Gamma (SRG) mission.  
After its launch from Baikonur, Kazachstan, SRG was put into a halo orbit around L2, whereupon it started
its all-sky survey in December 2019.  The eROSITA all-sky 
survey is carried out in a way very similar to the ROSAT all-sky survey, that is, the sky is scanned in great
circles perpendicular to the plane of the ecliptic.
The longitude of the scanned great circle moves by $\sim$1$^{\circ}$ per day, thus after half a year the whole sky is covered;
a detailed description of the eROSITA instrument and its hardware is presented by
\cite{2021predehl}.

For this study we only used the data from the first eROSITA all-sky survey, i.e., specifically
the first catalog covering the Western Galactic hemisphere published by
Merloni et al. (2024), who provide a detailed account of the eROSITA hardware as well as the
data products derived by the eSASS software system used for source detection and X-ray source catalogs creation;
it is important to keep in mind that
the quoted catalog count rates are fiducial count rates, in the sense that they refer to the on axis count rates
one would have measured if all seven eROSITA telescopes were observing simultaneously the source  under consideration
on axis.   During the survey operations the off-axis angle changes all the time, and not all seven telescopes 
may deliver useful data at any given time, hence the actually measured mean count rates are converted to
a fiducial on-axis count rate.
The SASS software performs source detection in various energy bands; in this paper we use the count rates 
derived from the so-called 1B detection in the 0.2~-2.3~keV energy band using a flux conversion 
factor of (ECF) of 9 $\times$ 10$^{-13}$ erg cm$^{-2}$ cnt$^{-1}$ to convert
count rates into energy fluxes. This ECF is appropriate for a thermal plasma emission with a temperature of 1~keV and an absorption column
of 3  $\times$ 10$^{19}$ cm$^{-2}$ and differs slightly from the fluxes quoted by Merloni et al. (2024), who used an ECF appropriate for
non-thermal sources.
No attempt was made to apply individual conversion factors to individual sources
based on their X-ray spectra.  Thus, the quoted X-ray fluxes refer to the energy band 0.2~keV~-~2.3~keV, which is 
well matched to the ROSAT band.  We have access to eROSITA data only at galactic longitudes
$l \ge $ 180$^{\circ}$, and furthermore, to avoid source confusion and any ensuing identification problems, we here consider only
the high galactic latitude sky with $|b| = 20^{\circ}$.  Finally we note that we used only sources with a detection likelihood $>$ 6;
Merloni et al. (2049) also published a supplementary catalog with sources with detection likelihood between 5 and 6, however, since the
level of contamination with spurious sources in this catalog is quite high we refrain from using this catalog for the time being.

\begin{figure}
\centering
\includegraphics[width=\hsize]{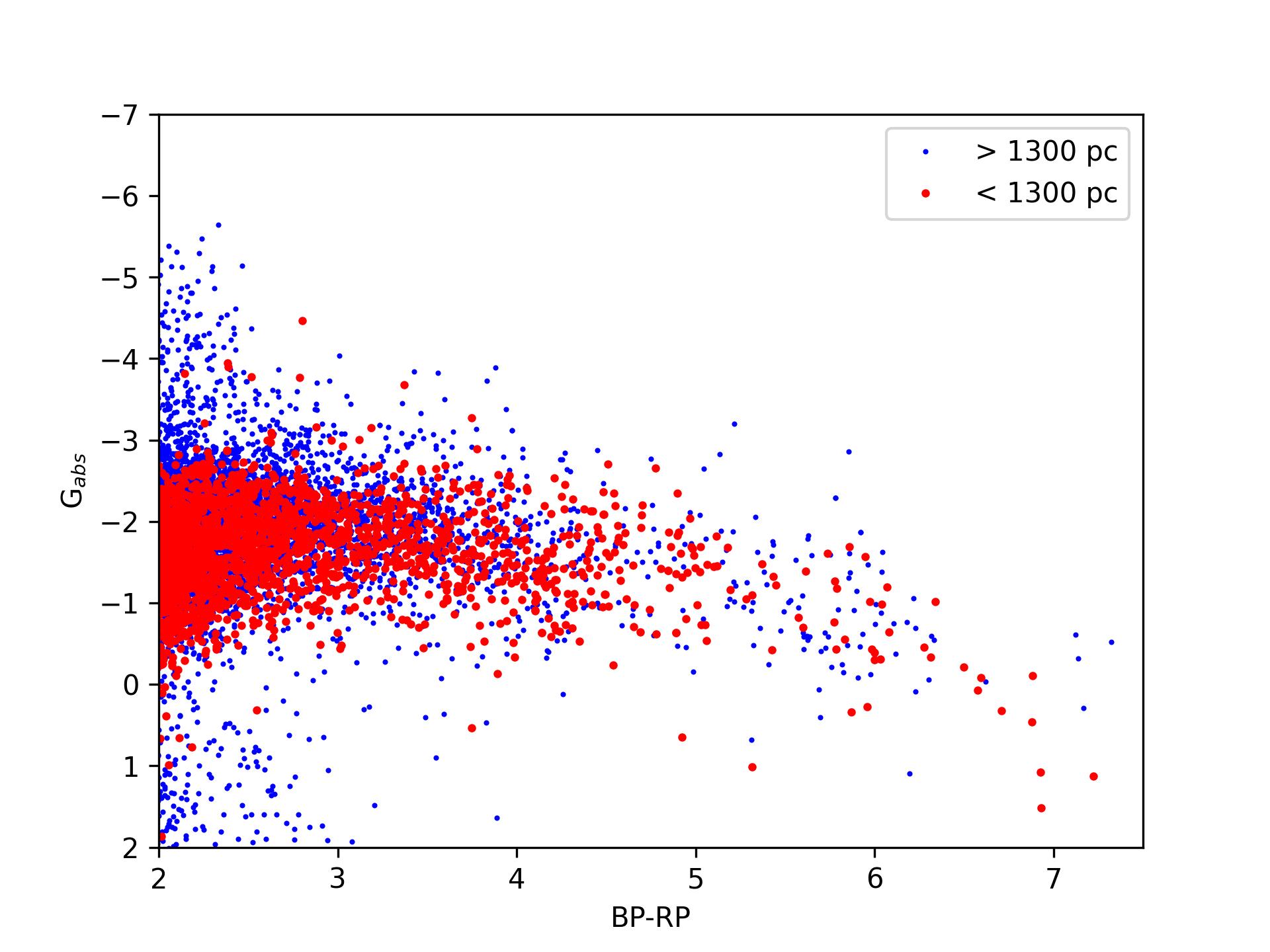}
\caption{Color-magnitude diagram diagram of high latitude red giant and supergiant stars; red symbols mark giants nearer than 1300~pc.}
\label{fig1}
\end{figure}

\subsection{Gaia data: Red giants and supergiants in the Milky Way}
\label{sec_redgiants}

Before discussing the eROSITA X-ray detected red supergiants, we examine the occurrence of such stars in
the {\it Gaia} catalog  in general. 
Red giants and supergiants are intrinsically very bright and can therefore be seen out to rather large distances, 
however, as is well known, the space density of those stars is quite low.
Using the new {\it Gaia} DR3 data we can actually construct a complete sample of such objects in the Milky Way.
If one considers only catalog entries with measured parallaxes and with absolute magnitudes brighter than 10
 (which corresponds to an early M-type star on the main sequence) and an apparent
magnitude brighter than 17.5 (otherwise an eROSITA  detection as a stellar X-ray source would be implausible), 
one  finds many millions of such entries.   However, if one restricts attention to red giants and supergiants, which are 
brighter $G_{abs}  $ ~=~ 0, and redder than the color BP-RP = 2, the corresponding catalog numbers are reduced very dramatically.   
To demonstrate this, we plot
a color-magnitude diagram (CMD) of such sources located in our X-ray search region 
at galactic latitudes larger than $|b| = 20^{\circ}$ and galactic longitudes $l > 180^{\circ}$ in  Fig.~\ref{fig1}.
Specifically focussing on the CMD region in the BP-RP range 2.5 $<$ BP-RP $<$ 7.5 and the $G_{abs} $  -7 $< G_{abs} < $ 2, we find only 7234 Gaia entries in this CMD region, out of which only 1735 are located closer than 1300~pc. Naturally these stars are concentrated towards the galactic plane, yet
it is clear that also in the optical such red giants and supergiants are rather rare objects, and we have constructed a complete sample of such objects in our study region in this fashion.

\begin{table}[t]
    \centering
    \caption{XMM-Newton observing details for Arcturus (Obs-Id 0865070101)\label{tab:xmm_arcturus}}
    \begin{tabular}{lcc}
        \hline
        \hline
        Detector & Configuration & Ontime (ks)\\
        \hline
        pn & Thick filter, Large Window  & 74 \\
        mos 1 & Thick filter, Large Window  & 81\\
        mos 2 & Thick filter, Large Window  & 81\\
        \hline
    \end{tabular}
\end{table}

\subsection{XMM-Newton data}

To obtain detailed spectral information on the corona from a low-activity giant, we observed
Arcturus for 80\,ks in continuous exposure with XMM-Newton; the details of this observation 
are listed in Tab.~\ref{tab:xmm_arcturus}.  The optical brightness of Arcturus required the
use of the ``thick filter'' for all EPIC detectors and mandates to switch off the optical 
monitor so that no simultaneous UV or optical data were obtained.
We processed the data with SAS version 19.1.0, using standard procedures and applying
standard filters; the observation was unaffected by background flares.

No strong X-ray source is
immediately visible at the nominal position of Arcturus. To check for possible weak X-ray emission
from Arcturus, we therefore concentrated on the pn-detector\footnote{Compared to the pn-detector,
the two MOS-detectors have a small effective area in our configuration. They show a 
weak nominal source photon excess in 
the same energy bands as the pn-detector, but with very low count numbers, i.e., 
contribute no significant additional information.} and experimented with source regions 
of different radii and different background regions (small and large circles, annuli, a 
combination of small circles, and a pie-shaped region, i.e., an annulus where the 
region immediately adjacent to the detector edge has been spared). These experiments 
do reveal a weak photon excess in the soft energy bands.  In the energy range between
300\,eV to 720\,eV range, which is centered on the O~{\sc vii} triplet,  one finds a photon 
excess between 16 and 19 photons (photons in source region - extrapolated background 
photons); the expected background in the source region is $\sim 50$ counts so that 
the smaller number corresponds to a 3.5\,\% statistical chance 
to be a background fluctuation while the larger one corresponds to  0.2\,\%. 
Therefore, we conclude that a small but statistically significant photon excess is detected at the 
position of Arcturus, while the resulting count rate itself, however, has an uncertainty of 
about 50\,\% due to the uncertainties in the background contribution.

\section{Results}
\label{sec_res}

\subsection{Color-magnitude diagram of stellar X-ray sources}
\label{colmag_sec}

\begin{figure}
\centering
\includegraphics[width=\hsize]{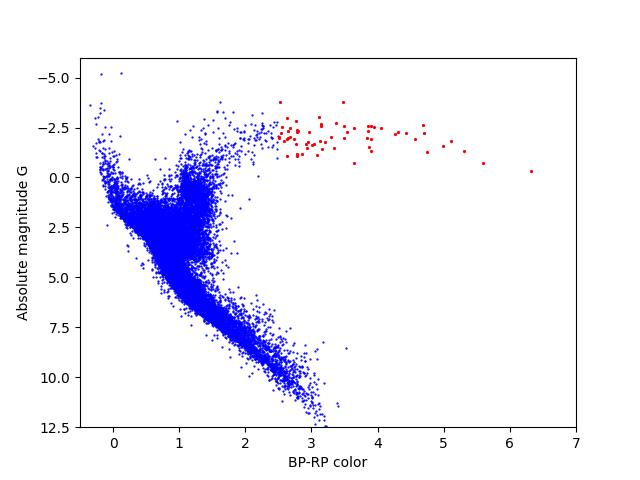}
\caption{Color-magnitude diagram of X-ray detected high latitude stars; red symbols mark giants with BP-RP $>$ 2.5;
Arcturus is expected to be located at BP-RP $\approx$ 1.55 and G $\approx$ -0.7.}
\label{colmag}
\end{figure}

In  Fig.~\ref{colmag} we plot  a color-magnitude diagram of high-latitude ($|b| > 20^{\circ}$)
and high probability (p$_{stellar}$ > 0.5, more details are provided in Sec.~\ref{xrayid}) of stellar eRASS1 detections.   As is obvious from
Fig.~\ref{colmag} and well-known from many previous studies,  stellar X-ray sources are found
all along the main-sequence and also in the regions where subgiants and giants are found.
For orientation, we estimated the BP - RP color and absolute G magnitude of Arcturus (BP - RP $\approx$ 1.55 and G $\approx$ -0.7),
Arcturus itself is not contained in Gaia DR3 because of its brightness.

An inspection of Fig.~\ref{colmag} reveals that there is a substantial number of stars that are more luminous and redder than Arcturus
itself.
Even more surprisingly one also encounters some sources among the very red giants and supergiants (here defined
as sources with counterparts redder than BP - BP = 2.5 and in the absolute magnitude range (-4 $<$ $G_{abs} < $ 0), marked 
with red symbols in Fig.~\ref{fig1}.  The overall number of these sources is rather small, and a list of these sources is provided  in Tab.~\ref{tab1},
giving there the relevant {\it Gaia} information as well as a flag indicating whether the X-ray counterpart is due to optical contamination
or a real X-ray source (cf., discussion in Sec.~\ref{opt_cont}).
It is very clear that genuine X-ray emission among those stars is quite unexpected and ought to be -- in all likelihood --
non-coronal. 

\subsection{Identification of X-ray emitting red giants and supergiants }
\label{xrayid}

In this section we discuss in detail the reliability of the identifications of the stellar X-ray sources shown in
Fig.~\ref{colmag}.
The eRASS1 catalog contains about 930000 point-like X-ray sources with galactic longitudes in excess of 
180$^{\circ}$.   To identify those X-ray sources that have red giants and supergiants as counterparts we followed
two different procedures.   We first followed the approach described by
\cite{2022schneider}, who aim at identifying {\bf all} coronal X-ray emitters in the eFEDS data;
for a detailed description of the mathematical basis of these identification procedures we refer to \cite{2023czesla}
and we did use the so-called HamStar catalog of coronal X-ray sources published by Freund et al. (2024).
In this approach each of eROSITA's catalogued X-ray sources is associated with
some probability $p_{stellar}$, i.e., probability that the counterpart of the X-ray source in question is a 
star.
Given our knowledge of the properties of stellar X-ray emission, it is clear that the sought for counterparts
must be listed as {\it Gaia} DR3 entries, and the probability $p_{stellar}$ is computed using various pieces
of information such as the offset between X-ray and optical position, the physical distance of the star and
the ratio between X-ray and bolometric flux of the counterpart; for details we refer to \cite{2022schneider}
and to Freund et al. (2024).
Since in this approach, (almost) the full totality of Gaia entries, which qualify as a potential stellar X-ray counterpart,
 is considered, one needs a training set to ``teach'' the algorithm about the properties of true coronal X-ray sources.

\begin{figure}
\centering
\includegraphics[width=\hsize]{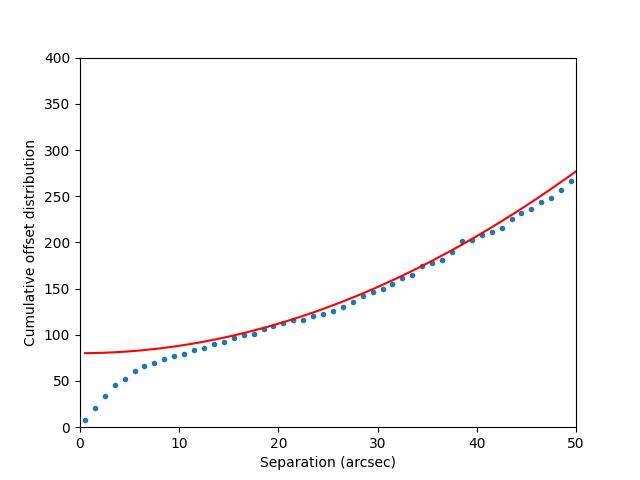}
\caption{Observed cumulative nearest neighbor distribution of red supergiants (blue dotted curve) together with model fit (red curve)
of the form A + B $e^{-C \rho ^2}$; see text for details.}
\label{fig2}
\end{figure}

Unfortunately, the training sets for red giants and supergiants are by necessity rather small, given the rarity of such objects both
at optical and X-ray wavelengths, and therefore the identification results may be less reliable than
for source classes for which much larger training sets are available.

However, since we have constructed a {\bf complete} sample of red giants and supergiants in the sky region of interest as
described in Sec.~\ref{sec_redgiants}, we actually do not have to consider the totality of all {\it Gaia} stars as possible counterparts, rather 
we can use the ``classical'' identification approach and simply look for straightforward positional coincidences only at the {\it a priori}
selected positions of these pre-selected objects, i.e., at the positions of the 7234 red giants and supergiants determined in Sec.~\ref{sec_redgiants}.
Matching then this target sample  with the eRASS1 source catalog,
we can construct the nearest neighbor distribution between the two samples; while clearly the nearest (optical) neighbor to an X-ray source
need not necessarily be the correct identification, for brighter stars this usually applies; for a rigorous mathematical
treatment of the position matching problem (also in source confusion situations) we again refer to \cite{2023czesla}.

Random matches may occur under all circumstances. For a uniform density distribution of such random ``background'' sources 
$\eta_b$, the distance between the nearest such ``background'' source and some specified position is given by the
nearest neighbor distribution. Hence the (cumulative) probability P$_{cum}( \rho)$ of finding one or more random sources within the distance 
$\rho$ is given by the expression $P_{cum}(\rho) = 1 - e^{-\pi \eta_b \rho^2}$.
To obtain the number of true matches $N_{true}$ among the $N_{tot}$ considered positions, we then
fit the observed cumulative distribution
$N_{cum,obs}$ to the expression $N_{true}$ + ($N_{tot}$ - $N_{true}$) $\times P_{cum} (\rho)$ 
to compute the true number $N_{true}$ of
non-random matches.  The result of this exercise is shown in Fig.~\ref{fig2}, and from the offset we may estimate
$N_{true} \approx 80 \pm 5$ as the estimated number of true (rather than random) matches.

\begin{figure}
\centering
\includegraphics[width=\hsize]{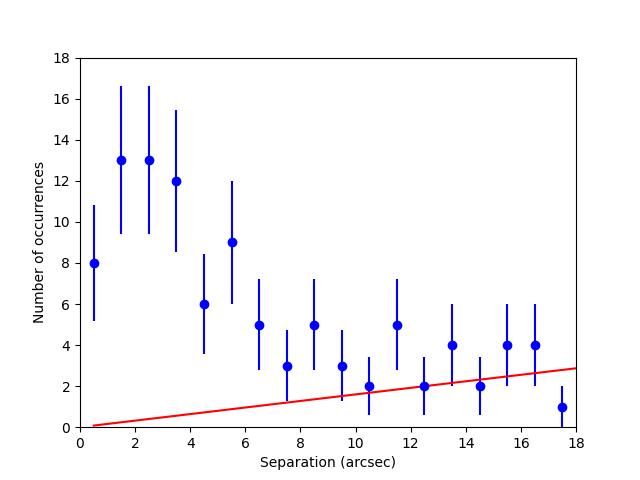}
\caption{Differential nearest neighbor distribution of red giants and supergiants (blue dots) as a function of position offset and modeled contribution of
spurious identifications  (red solid curve); see text for details.}
\label{fig3}
\end{figure}

In Fig.~\ref{fig3} we examine in detail the measured differential nearest neighbor distribution vs. the expected model distribution; it is obvious that for offset distances
below 6 arcsec a clear excess is observed, and an excess may extend up to 10 arcsec before it merges with the random background sources.   To be on the safe 
side and capture all possible counterparts, we first
consider all matches with offsets below 12.5~arcsec between X-ray and optical positions as potential counterparts and list these 96 objects in Tab.~\ref{tab1};
the columns in Tab.~\ref{tab1} refer to the {\it Gaia} DR3 numbers,
right ascension and declination as well as {\it Gaia} G band magnitude and BP-RP color (columns 1-5), the {\it Gaia} distance computed from the
inverse parallax (column 6), the matching distance between the {\it Gaia} position and the eRASS1 X-ray position (column 7), the stellar probability
(column 8), a flag indicating optical contamination (column 9, see section \ref{opt_cont} for a detailed discussion) and an asterisk
in column 10, denoting the ``good'' candidates.  The stellar probability listed in column 8 of Tab.~\ref{tab1} denotes the stellar probability computed from
our ``general'' identification scheme.   Inspection of Tab.~\ref{tab1} shows that these probabilities become small for large offsets (as naturally expected),
and furthermore, the majority of our accepted candidates (marked with asteriks in Tab.~\ref{tab1}) have stellar probabilities larger than 0.5 with three
exceptions.

\begin{figure}
\centering
\includegraphics[width=\hsize]{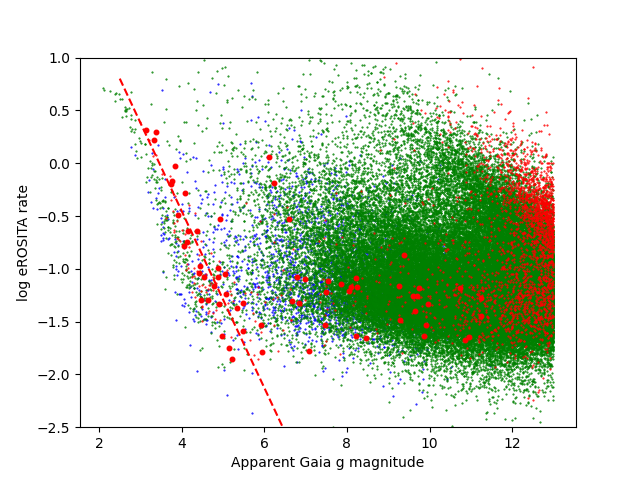}
\caption{eROSITA count rate vs. apparent {\it Gaia} G magnitude for stellar X-ray sources; blue data points
refer to stars with BP-RP $<$ 0.25, green data points to stars with  0.25 $<$ BP-RP $<$ 2. and (small) red 
data points to stars with BP-RP $>$ 2; the large red dots refer to stars considered in this paper.}
\label{fig5}
\end{figure}

\begin{table*}[h]
\tiny
\begin{tabular}{  | r  |  r  |  r  |  r  | r  | r  |  r  |  r |  r  |  r  | r  |  r  | }
\hline
Gaia DR3 ID & Name & BP-RP & G$_{abs}$ & Spec & CR & Err & Exp. & Li & f$_{X,-14} $& log(L$_X$)&FUV - NUV\\
\hline
4644856055150677632 & CI Hyi & 4.31 & -2.27 & M5/7III & 0.083 & 0.023 & 217.6  & 26.8    & 7.50 & 30.69 & n.a.\\
4748477123328077696 & R Hor & 5.60 & -0.70 & M5-7e & 1.143 & 0.072 & 224.4   & 1447.2 & 102.86$^*$ & 31.38 $^*$& n.a.\\
5159194331764625920 & DR Eri & 2.64 & -1.96 & M2/3III & 0.030 & 0.014 & 191.0 & 7.8      & 2.66 & 30.31 & 0.29\\
5100253911446411264 & BH Eri & 3.86 & -2.34 & M6 & 0.082 & 0.023 & 195.2       & 27.5     & 7.40 & 30.98& n.a.\\     
4723045075781077760 & UU Ret & 2.64 & -2.31 & M3III & 0.023 & 0.009 & 400.2  & 13.4 & 2.10 & 30.43 & n.a.\\  
4616109220563904768 & U Men & 3.75 & 0.53 & M3e & 0.647 & 0.052 & 266.4     & 637.2 & 58.24$^*$ & 31.28$^*$ & 2.35\\
4772353701224784640 & CD-51 1503 & 2.96 & -1.90 & S5,2 & 0.113 & 0.017 & 451.0 & 105.3 & 10.19 & 31.08& n.a.\\ 
4620697620025724544 & CPD-8486 & 3.01 & -1.64 & S4,4 & 0.071 & 0.017 & 314.5  & 41.6 & 6.43 & 30.71 & 2.59\\   
893132234387415808  & BD+32 1528 & 3.08 & -1.12 & & 0.068 & 0.032 & 77.0          & 11.0 & 6.14 & 30.87& 1.91 \\ 
5193680960648176768 & TYC9506-1836-1 & 3.79 & -1.52 & & 0.063 & 0.015 & 357.2 & 41.7 & 5.63 & 30.76& 3.13 \\
5761301055347596800 & MV Hya & 2.63 & -1.95 & M3/4III & 0.050 & 0.026 & 87.8     & 9.2   & 4.50 & 30.38 & n.a.\\
645414291872801536  & EH Leo & 2.76 & -2.84 & & 0.296 & 0.066 & 75.2                  & 71.1 & 26.68 & 31.31 & 2.09\\
5195956296883496832 & HD 84048 & 2.58 & -1.84 & M1/2 & 0.067 & 0.014 & 396.9 & 56.1 & 6.04 & 30.80 & 0.41\\
794754943320201600  & R LMi  & 6.31 & -0.33 & M6.5-9e & 0.079 & 0.033 & 79.7    & 18.0 & 7.15$^*$ & 30.32$^*$& n.a.\\
3491583497096278144 & WW Crt & 2.85 & -1.18 & M4III & 0.653 & 0.076 & 128.1    & 287.9 & 58.78 & 31.25 & 3.35\\
6173329362681684736 & TW Cen & 5.30 & -1.34 & M6-8Ib-II & 0.067 & 0.022 & 172.0 & 18.5 & 6.06 & 30.67 & 3.99\\
6738529604088911104 & IRAS19204-3959 & 3.16 & -0.70 & & 0.061 & 0.032 & 82.2     & 8.9 & 5.52 & 30.82 & n.a.\\
6713213967334901376 & HD 181817 & 2.51 & -2.04 & M2III & 0.061 & 0.029 & 83.7     & 10.3 & 5.47 & 30.65 & 2.68\\
6342796540114659200 & HD 199203 & 2.56 & -2.55 & M2/3(III) & 0.076 & 0.024 & 177.4 & 20.9 & 6.86 & 30.86 & n.a.\\
\hline
\end{tabular}
\vskip 0.25cm
\caption{\label{tab2} Physical and X-ray properties of very red giant stars identified with X-ray sources. Columns 3 and 4 give the HR diagram position, i.e. the 
Gaia BP-RP colour and absolute G magnitude, col. 6 and 7 list the count rates per second and the corresponding errors, col. 8 the exposure time in seconds, col. 9 the 
likelihood, col. 10 the X-ray flux in $10^{-14}$ erg/s, col. 11 gives the derived X-ray luminosities in erg/s, and the last column gives the FUV - NUV colors derived from GALEX; see text for details.
The stars marked with "*" are likely affected by optical contamination;
see again text for details.  }
\end{table*}

\subsection{Optical contamination}

\subsubsection{The overall picture}
\label{opt_cont}
eROSITA data is known to be subject to optical contamination; since M-giants can be very bright optically, it behooves us well to examine the influence
of optical contamination on our results.  For example, the two very bright red supergiants of type M, $\alpha$ Ori ( V = 0.42) and $\alpha$ Sco ( V = 0.91),  are actually
not listed in the {\it Gaia} catalog because of their immense optical brightness, yet a visual inspection shows these two stars to be contained
in the eROSITA catalogs although no X-ray emission from these stars has been detected in the ROSAT all-sky survey.  
In more general terms, we can plot -- in~Fig.~\ref{fig5}~-- 
the measured (logarithmic) eROSITA count rate vs. the optical G magnitude of the counter parts for objects with counterparts
brighter than about g $\approx$ 13.   Fig.~\ref{fig5}  shows that the bright end 
of the distribution is more or less linear, where the parameter of the regression curve depend somewhat on stellar color.   The regression plotted
in Fig.~\ref{fig5}  has been derived from our sample stars with apparent g magnitudes brighter than 5.75 and is given by the expression
\begin{equation}
log(Cr_{X-ray}) = 3.05 - 0.88 \times  g, 
\label{opt_cont_eq}
\end{equation}
where Cr$_{X-ray}$ denotes the eROSITA X-ray count rate caused by optical contamination and $g$ the apparent Gaia g magnitude.

The linearity between optical and X-ray flux for the brighter Gaia stars
suggests optical contamination and not true X-ray flux to be the cause for the eROSITA detections in the vicinity of the
dashed line in Fig.~\ref{fig2}; these sources are therefore marked by ``Y'' in Tab.~\ref{tab1}.   On the other hand, there are also quite a few sources located far away from the optical contamination
line in Fig.~\ref{fig5}, and these sources are clearly true X-ray sources marked by ``N'' in Tab.~\ref{tab1} , and we focus upon these sources in the following.

\subsubsection{The Mira stars R Hor, R LMi, U Men, and TW Cen}
\label{opt_cont_ind}

\begin{figure}
\centering
\includegraphics[width=\hsize]{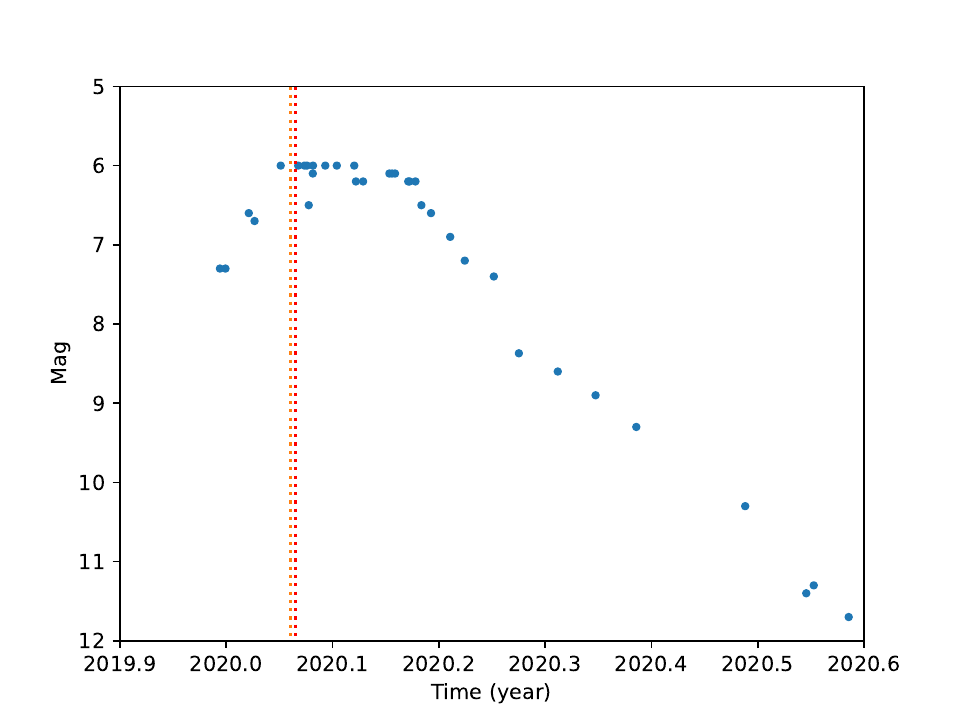}
\caption{Optical light curve (derived from visual V band observations, with data taken from AAVSO data base) of R~Hor (blue data  points) together with scanning time in eRASS1 (red dotted vertical lines).}
\label{aavso_rhor}
\end{figure}

In our sample there are some Mira-type variable stars which can change their optical flux by a couple of magnitudes, and the magnitude listed
in the {\it Gaia} catalog must not necessarily represent the actual magnitude  at the time of the eRASS1 observations.  One such case is the Mira-like
variable R~Hor.   Using the data from the AAVSO data base, we can construct  the optical light curve of R~Hor (shown in Fig.~\ref{aavso_rhor})  and estimate
its brightness during the time of the eRASS1 observations.   As is obvious from  Fig.~\ref{aavso_rhor}, the eRASS1 observations took place during the maximal
brightness of R~Hor at a level of about 6 mag, yet this magnitude corresponds to the visual band.   
Using Tab. 5.7 in the Gaia EDR3 user guide and the very red BP-RP color of 5.6 we compute an estimated G-V color of -4.2, which leads to an apparent Gaia magnitude of g = 1.8. From our regression equation (1) the observed eROSITA count rate of 1.14~s$^{-1}$ would result in a g magnitude of about 3.4 if related solely to optical contamination. While these two estimates
do not agree (note that the color is likely to change during a pulsation cycle), it is however clear that also the apparent g magnitude of R~Hor was quite large and
thus we conclude that the X-ray detection of R~Hor during eRASS1 is clearly due to optical contamination.

\begin{figure}
\centering
\includegraphics[width=\hsize]{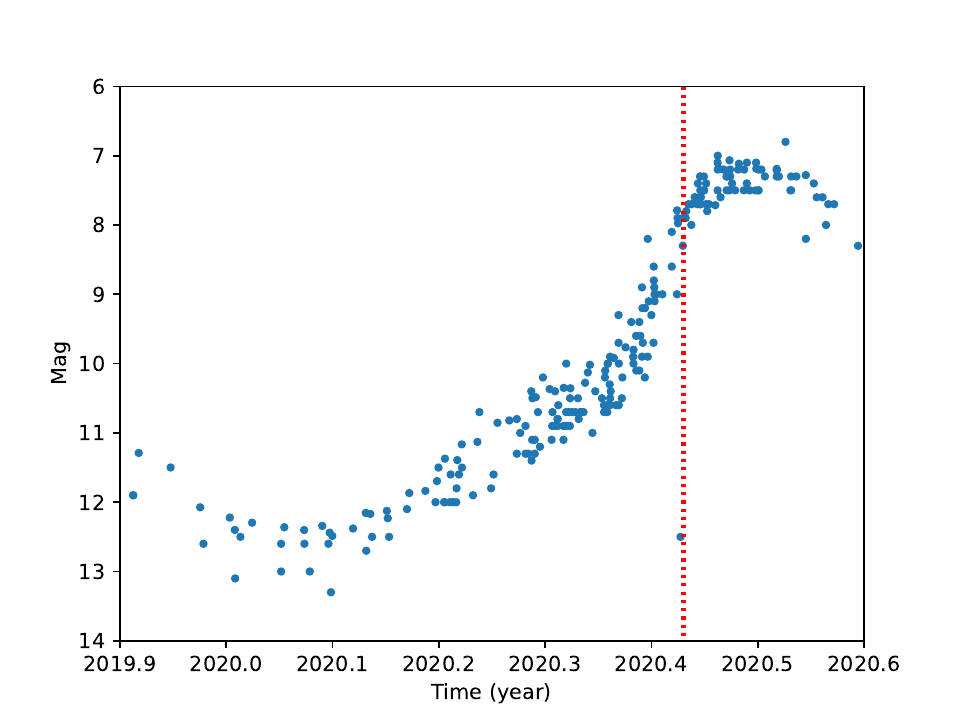}
\caption{Optical light curve (derived from visual V band observations, data taken from AAVSO data base) of R~LMi (blue data  points) together with scanning time in eRASS1 (red dotted vertical lines).}
\label{aavso_rlmi}
\end{figure}

A similar case is R~LMi which is also a Mira-like variable, the light curve of which is shown in Fig.~\ref{aavso_rlmi}.   The light curve
shows that the eRASS1 observations took place about a month prior to the optical maximum of R~LMi and that
its magnitude was at that time was around (visual) mag 8.   Thus the (visual) magnitude difference between R~Hor and R~LMi at the times of their 
respective eRASS1 observations was about 2 with substantial error given the noise in the light curve data (cf., Fig.~\ref{aavso_rhor} and Fig.~\ref{aavso_rlmi}), 
corresponding to a flux ratio of 0.16 if both detections are assumed to be due to optical contamination.
From Tab.~\ref{tab2} we compute an observed ratio of 0.07, i.e., a factor two within ``expectations'', thus we conclude that the X-ray detection
of R~LMi is again very likely due to optical contamination.

For the star U Men there are unfortunately no optical data available in the AAVSO database contemporaneous
to the eRASS1 observations. However, from the light curve resulting from previous observations and considering the well-known 
407~d period of U Men, we estimate that the eRASS1 observations fell into a phase close to maximum brightness of U Men, which is typically V $\approx$ 8. Hence, as in the case of R LMi and to be on the safe side, we also consider U Men to be optically contaminated.  However, 
the situation is different for TW Cen, where two data points available in January 2020 suggest a visual magnitude not brighter than 12 at the time of the eRASS1 observations. We therefore assume TW Cen not to be optically contaminated.

\subsection{Sample selection}
\label{sample}

In order to arrive at a reasonably clean sample we need to restrict the number of stars listed in Tab.~\ref{tab1}.  While, clearly, some X-ray emission
might be hidden for the stars marked with the optical contamination flag in Tab.~\ref{tab1}, we cannot distinguish X-ray and optical signals,
therefore those stars are excluded from further consideration.  Next, inspection of the distance distribution shows a minimum near
1300~pc; we therefore impose an arbitrary cut at this distance, but do point out that there are quite a few true X-ray emitting red giants and supergiants
beyond this distance limit and usually there is only rather little additional information on these sources.  Finally, we reject all stars with matching distances
larger than 7~arcsec or a stellar probability less than 0.5.   In this fashion we arrive at a ``clean'' sample of 23 stars, from which we exclude the
sources R~Hor, R~LMi, and U~Men, since we attribute the detections of these Mira like stars to optical contamination as discussed in detail
in Sec.~\ref{opt_cont_ind}.
This ``clean'' sample (as well as the data for the three Mira stars with optical contamination)  is listed in Tab.~\ref{tab2}, where we
provide, again, the Gaia DR 3 numbers of the sample stars, an additional name whenever available, spectral type and the relevant X-ray information.
We specifically point out that the detection likelihoods of these sources are quite large, only a rather smaller number of sources has a
detection likelihood below 10, where one might start quibbling about the reality of these sources.

\subsection{Spectral analysis of the XMM-Newton Arcturus spectrum}
\label{arc_spec}

In this section we investigate whether Arcturus can be considered a ``prototypical'' X-ray emitting giant. 
Unfortunately, the rather small number of recorded source photons from Arcturus prohibits a detailed spectral modeling
of the X-ray data.  Nevertheless, the number of source photons recorded in different energy bands does allow some estimates
of the plasma temperature.  For the following analysis we use a pie-shaped background region, 
since we expect this choice to lead to the most realistic background 
estimate; furthermore,  the resulting numbers are generally intermediate for other choices of 
background regions. 

An investigation of the spectral distribution of the source photons 
shows that most 
photons are detected between 300 and 720\,eV, a marginal 
excess may also be present at energies between 200 and 300\,eV but is not significant
and rather confirms that optical contamination is insignificant for our XMM-Newton data. 
Also, no significant
excess is seen at energies above 720\,eV with the measured counts matching expectations from
background extrapolations well, i.e., only soft source photons are detected. This pattern 
is indicative of emission from a (very) cool plasma compared to  stellar coronae; we further note
that it is the use of the thick filter with its small effective area that leads to the small number of
detected photons below 300\,eV. 
The lack of more energetic source photons does imply a plasma temperature below 3\,MK; 
otherwise, the predicted number of 725--1500\,eV source photons would violate 
the 90\,\% upper limit. The nominally measured band count rates are best compatible with 
a plasma temperature around 1.5\,MK, leading  a 0.2-2.0\,keV flux of 
about $1.3\times10^{-15}$\,erg\,s$^{-1}$\,cm$^{-2}$ ($L_X = 2\times10^{25}$\,erg\,s$^{-1}$). 
Requiring that Arcturus
has the same nominal flux as reported by \citet{Ayres_2018}, would require a plasma 
temperature closer to 1\,MK while temperatures below $8\times10^5\,$K would produce
an equal number of 200--300\,eV and 300--725\,eV photons, incompatible with the data.
In summary, the XMM-Newton data of Arcturus are compatible with the Chandra HRC-I 
data for plasma temperatures around 1\,MK and independently restrict the plasma
temperature to log\,T = 5.9\dots6.5. Within this temperature range, the  
energy conversion factor (ECF) may deviate by approximately 50\,\% from its value at 
the nominal temperature. 

We do point out that the above flux and temperature estimates ignore any potential circumstellar absorption, and
the data are surprisingly insensitive to adding moderate amounts of absorption. For example, a column density
of $N_H = 2\times10^{21}$\,cm$^{-2}$ shifts the corresponding nominal plasma temperature to
1\,MK, but remains compatible with the data. This is due to two factors; first, 
increasing the absorbing column density can be compensated by decreasing the plasma temperature to some degree.
And second, the rather high background in the lowest energy bands prevents strong flux constraints 
w.r.t. intermediate photon energies (above 0.3\,keV), i.e., an observed spectrum without 
flux below 0.3\,keV is compatible with the data. For the Arcturus XMM-Newton data, 
column densities above 
$5\times10^{21}$\,cm$^{-2}$ start to produce too few photons below 500\,eV even for very low 
plasma temperatures of 0.5\,MK; the latter case corresponds to an emitted flux of $10^{-12}$\,erg\,s$^{-1}$\,cm$^{-2}$
and an intrinsic X-ray luminosity $L_X$ of $1.5\times10^{28}$\,erg\,s$^{-1}$.
However, whether such low ``coronal'' temperatures\footnote{Often, the corona is defined as the region with plasma temperatures in excess of 1\,MK.} are realistic is debatable.
At any rate, requiring that the plasma temperature is $>10^6\,$K, implies a column density 
below $2\times10^{21}$\,cm$^{-2}$.  

The nominal X-ray flux from Arcturus of about  $1.3\times10^{-15}$\,erg\,s$^{-1}$\,cm$^{-2}$
is somewhat lower than the value reported by 
\citet{Ayres_2018} from combined Chandra HRC-I data ($2\times10^{-15}$\,erg\,s$^{-1}$\,cm$^{-2}$), but compatible 
within the uncertainties. Interestingly, \citet{Ayres_2018} noted that the count rate for the 2018 Chandra observation is 
about twice as high as during 2002, with large uncertainties however. Thus, the somewhat lower nominal flux during
the XMM-Newton observation may indicate some variability in the X-ray flux of Arcturus. In any case, Arcturus is
surely among the X-ray faintest objects detected so far, and its
X-ray emission does stem from rather cool plasma around 1\,MK; there is no evidence for 
significant absorption, yet the limits on the absorbing column density, however, permit moderate
absorption columns of up  to  $2\times10^{21}$\,cm$^{-2}$.

\begin{figure}
\centering
\includegraphics[width=\hsize]{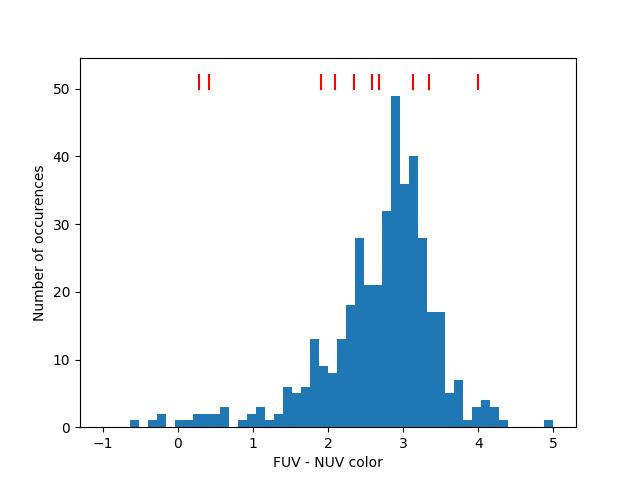}
\caption{Distribution of FUV - NUV colors derived by GALEX for a sample of nearby (d $<$ 1300~pc) M supergiants; the red 
marks at the top indicate the GALEX FUV - NUV colors (whenever available) for our sample of X-ray detected supergiants
(see text for details).}
\label{galex_col}
\end{figure}

\section{Discussion and conclusions}
\label{sec_discuss}

In this paper we present a systematic study of X-ray emitting red giants and supergiants in our Galaxy
using the new X-ray data from the first of eROSITA's all-sky surveys.    Using the
photometric and astrometric data from the {\it Gaia}, we can construct a complete sample
of such objects in the western galactic hemisphere, in which we can search for intrinsic X-ray emission.

Our pilot study of X-ray emission of M-type giants and supergiants with data from the first of
eROSITA's all-sky surveys has revealed a number of true and genuine X-ray emitters.
By restricting attention to high galactic latitudes, we have significantly reduced the
problem of erroneous positional misidentifications.   A latent problem with all eROSITA data
is optical contamination, which may affect the brighter M giant population.  Furthermore,
Mira-like variables may change their optical brightness substantially, hence it is
important to check their optical brightness at the time of the X-ray observations;  in this
fashion we attribute the ostensible X-ray detections of R~Hor, U~Men and R~LMi 
to optical contamination.

The X-ray luminosities of the securely detected M-giants are concentrated in a comparatively  narrow
range between 10$^{30}$ erg s$^{-1}$ and 10$^{31}$ erg s$^{-1}$.  Since the optically brighter objects
of this kind are typically optically contaminated, unfortunately, many of the
counterparts identified by us have been little studied so far.   As described already in section \ref{sec_intro},
one does not expect M-type giants to be intrinsic X-ray emitters.   Since the number of spurious identifications
is low, one cannot attribute the identified counterparts simply as spurious coincidences, rather
one expects hitherto unseen companions to be the origin of the observed X-ray emission, 
yet the number of {\bf known} binaries in this sample is rather low.

The new XMM-Newton data for Arcturus show that the X-rays detected from the position of Arcturus are predominantly soft with 
the majority at energies below 0.7\,keV. 
Extrapolating the Arcturus results suggests that the fate of X-ray emission near the dividing line is 
caused by a true decrease 
in coronal emission, not just extinction of bright X-ray emission of an intrinsically ``powerful'' corona. 
If Arcturus is indeed typical for the giant stars near the XDL, the results imply that
the X-ray flux {\it and} the plasma temperature decrease when moving from left to right of the dividing line, 
and both features together make stars to the right of the dividing line X-ray faint.

The XDL is located in the HRD close to the region where giants develop strong winds \citep{Reimers_1977},
which may channel part of the energy away from the corona towards these winds. Furthermore, the large scale height
of giants to the right of the XDL permits static cool loops \citep[$T\sim10^5\,$K, see][]{Antiochos_1986}, and this 
may be observed for Arcturus, i.e., large but cool loops; and because the X-ray luminosity scales 
with plasma temperature $T$ as $T^{3.5}$ \citep[see summary in][]{Guedel_2004}, the total X-ray luminosity 
of Arcturus is as low as observed with $L_X/L_{bol}\sim10^{-11}$ without needing to resort to absorption 
effects. 

These results clearly show that the example of Arcturus cannot be prototypical for the X-ray emission from the red supergiants
detected by eROSITA. How should one then interpret  observed X-ray emission of these red supergiants ?   
Given the X-ray luminosities found for these objects, active late-type stars would indeed constitute potential candidates.
However, if this were true, these secondaries would have to be fairly young and consequently
the M-type primaries in these systems would also have to be young and hence relatively massive.
Yet an inspection of, for example, PARSEC isochrones (cf., Fig.~17 in \cite{2022nguyen}) shows
that such primaries would have G magnitudes larger than those observed for our sample stars (as shown
in Fig.~\ref{colmag}), and we conclude that neither the hypothesized primaries nor secondaries are particularly young.
We therefore rule out active stars as possible companions and must thus look for counter parts among older stars.

In this context we recall the star 56~Peg (=~HR~8796~=~HD~218356),  a rather bright, mild barium star of spectral type K0.5II.
56~Peg was first detected as an X-ray source by \cite{1982schindler} using the {\it Einstein Observatory}, and this
detection was confirmed by \cite{1998huenscha} using data from the ROSAT all-sky survey. 56~Peg also appears
in the XMM-Newton slew survey catalog \citep{2008saxton}, hence there can be no doubt that
56~Peg is a persistent X-ray source.   \cite{1982schindler}  report an X-ray luminosity of 3 $\times$ 10$^{31}$ erg/s,
which is reduced by 30\% if one adopts the {\it Gaia} distance of 181~pc rather than the distance adopted by \cite{1982schindler} at the time;
thus the X-ray luminosity of 56~Peg is quite consistent with the X-ray luminosities of our sample stars.

While 56~Peg is much ``bluer'' than the giants studied in this paper (it is of spectral type K0.5~II), the arguments put forward
by \cite{1982schindler} can be applied in very much the same fashion. 
In the case of 56~Peg, \cite{1982schindler}  argue that the observed X-ray emission cannot be coronal, rather they conclude
that the observed X-ray emission stems from a companion white dwarf, which at the time was only known from UV
spectroscopy.  In the mean time extensive radial velocity work was carried out which shows that 56~Peg is indeed a spectroscopic
binary if not a triple system; \cite{2006griffin} reports a period of 111~days, \cite{2023escorza} even report a spectroscopic triple
with periods of 111~days and about 15000~days, with the latter period belonging to the more massive companion.
Therefore the white dwarf postulated on the basis of ultraviolet spectroscopy and X-ray emission seems to have been found
through extensive long-term radial velocity work.

While we do not have UV spectroscopy for our sample stars, we examined the results of the all-sky imaging survey (AIS) performed
by the GALEX satellite \citep{bianchi2017} in a NUV band with $\lambda_{eff} \sim $ 2310~\AA \ and in a FUV band with
$\lambda_{eff} \sim $ 1528~\AA .  In Fig.~\ref{galex_col} we provide a FUV-NUV color histogram for those of our (complete) sample stars
located closer than 1300~pc and with valid FUV {\bf and} NUV measurements.   While it is difficult to interpret the observed
color distribution because of extinction effects, it is clear that the observed distribution is concentrated in the color range
2 - 3.5 with some ``tail'' of the observed color distribution below values of unity.   In Fig.~\ref{galex_col} we also indicated the 
observed GALEX  FUV - NUV colors (whenever available) for the X-ray detected supergiants (see also Tab.~\ref{tab2}).
As is apparent from Fig.~\ref{galex_col}, most of these values are located in the bulk of the distribution, with two
exceptions, namely the stars DR~Eri and HD~84048, which are much ``bluer'' than the bulk of the observed population; 
we note in this context that the observed FUV-NUV color of 56~Peg is 1.12, i.e., also in the ``blue'' tail of the FUV-NUV color distribution despite the fact
that the primary is of spectral type K0II, i.e., much earlier than our sample stars of spectral type M.  
Thus the stars DR~Eri and HD~84048 are obviously the prime candidates for the
presence of ``hidden'' WD companions, while for the other sample stars there is no evidence for hot companions 
at least from the available GALEX UV data.

\cite{1982schindler} propose wind accretion of a small fraction of the emanating wind onto the surface of a white dwarf as the
energy source for 56~Peg and argue that by assuming reasonable mass loss and accretion rates
this mechanism can explain the observed X-ray emission.  Given that also the giants investigated in this paper
have strong winds the same mechanism should be operating if indeed the X-ray sources reported in this paper
{\bf all} have white dwarf companions.    Finding these hypothesized companions is challenging and requires
more sensitive UV data and/or extensive radial-velocity monitoring, which, however, 
could be carried out with rather modest-sized telescopes.   If our hypothesis is correct,
these stars would show some resemblance to the class of barium stars, which were shown, most recently by \cite{2019jorissen},
to be all binaries.  Another possibility would be to search for chemical peculiarities in the present-day primaries other than barium;
as described in detail for example by \cite{2016merle}, the hypothesized present-day white dwarf must have been located at some point
in the past on the asymptotic giant branch producing s-process elements, which might still be visible as pollutants in the spectra
of the then secondary and present-day primary.   Thus in summary, the puzzle of X-ray emitting ``forbidden'' M-type supergiants
is not yet solved, however, with the stellar X-ray sources presented in this paper a sufficiently large sample has become available
which has the potential to actually solve the puzzle.

\begin{acknowledgements}
 
This work is based on data from eROSITA, the primary instrument aboard SRG, a joint
Russian-German science mission supported by the Russian Space Agency
(Roskosmos), in the interests of the Russian Academy of Sciences 
represented by its Space Research Institute (IKI), 
and the Deutsches Zentrum f\"ur Luft- und Raumfahrt (DLR). 
The SRG spacecraft was built by Lavochkin Association (NPOL) 
and its subcontractors, and is operated by NPOL with support from
IKI and the Max Planck Institute for Extraterrestrial Physics (MPE). 
The development and construction of the eROSITA X-ray instrument was
led by MPE, with contributions from the Dr.\ Karl Remeis Observatory 
Bamberg \& ECAP (FAU Erlangen-N\"urnberg), the University of Hamburg 
Observatory, the Leibniz Institute for Astrophysics Potsdam (AIP), 
and the Institute for Astronomy and Astrophysics of the University 
of T\"ubingen, with the support of DLR and the Max Planck Society. 
The Argelander Institute for Astronomy of the University of Bonn and 
the Ludwig Maximilians Universit\"at M\"unchen also participated in the science 
preparation for eROSITA.
The eROSITA data used for this paper were
processed using the eSASS/NRTA software system developed by the
German eROSITA consortium. 
This work has made use of data from the European Space Agency (ESA)
mission {\it Gaia} (\url{https://www.cosmos.esa.int/gaia}), processed by
the {\it Gaia} Data Processing and Analysis Consortium (DPAC,
\url{https://www.cosmos.esa.int/web/gaia/dpac/consortium}). Funding
for the DPAC has been provided by national institutions, in particular
the institutions participating in the {\it Gaia} Multilateral Agreement.
We gratefully acknowledge the variable star observations from the AAVSO 
International Database contributed by observers worldwide and used in this 
research.  Furthermore,
this research has also made use of the SIMBAD database,
operated at CDS, Strasbourg, France.

\end{acknowledgements}

\bibliographystyle{aa}
\bibliography{bibfile}

\begin{appendix}

\section{Notes on individual detected SRG sources}
\label{sec_individual}

In the following section we put together some notes on individual objects:\\
\\{\it CI~Hyi}:
The only object in our sample with previously reported X-ray emission:
\cite{2015sahai} discuss the X-ray emission of CI~Hyi in their pilot study of 
AGB stars with far-ultraviolet excesses (i.e., so-called fuvAGB stars) using data from the XMM-Newton and {\it Chandra} Observatories and find that the X-ray emission is variable on a time scale of hours;
CI~Hyi is a long period variable of the type SRB type.\\ 
{\it DR Eri}: LPV candidate of  type LB with a period of 51.3~d\\
{\it BH Eri}: LPV, type SR; period 89,9 d\\
{\it UU Ret}: LPV, type SRB\\
{\it U Men}: OH/IR star, type M, period 409 d; this is a binary (Baize 1962, Proust et al. 1981)\\
{\it MV Hya}: LPV, type SRB\\
{\it EH Leo}: LPV, type SRB, period 39,34 d\\
{\it WW Crt}: LPV, type SR\\
{\it TW Cen}: Mira star, period 271~d\\

\end{appendix}


\end{document}